\documentclass[aps,prl,twocolumn,showpacs,referee,superscriptaddress,showpacs, amsmath, amssymb]{revtex4}

\usepackage{graphicx}
\usepackage{dcolumn}
\usepackage{bm}

\begin{document}

\def\btt#1{{\tt$\backslash$#1}}
\def\BibTeX{\rm B{\sc ib}\TeX}
\draft

\title{Hydromagnetic Instability in Differentially Rotating Flows} 
\author{Alfio Bonanno}
\affiliation{ INAF, Osservatorio Astrofisico di Catania,
           Via S.Sofia 78, 95123 Catania, Italy} 
\affiliation{INFN, Sezione di Catania, Via S.Sofia 72,
           95123 Catania, Italy} 
\author{Vadim Urpin}
\affiliation{ INAF, Osservatorio Astrofisico di Catania,
           Via S.Sofia 78, 95123 Catania, Italy} 
\affiliation{A.F.Ioffe Institute of Physics and Technology, 
           194021 St. Petersburg, Russia}

\date{\today}
\begin{abstract}
We study the stability of a compressible  differentially rotating 
flows in the presence of the magnetic field, and we show that the 
compressibility profoundly alters the previous results for a magnetized 
incompressible flow. The necessary condition of newly found instability 
can be easily satisfied in various flows in laboratory and astrophysical 
conditions and reads $B_{s} B_{\varphi} \Omega' \neq 0$ where $B_{s}$ and 
$B_{\varphi}$ are the radial and azimuthal components of the magnetic 
field, $\Omega' = d \Omega/ds$ with $s$ being the cylindrical radius. 
Contrary to the well-known magnetorotational instability that occurs 
only if $\Omega$ decreases with $s$, the instability considered in this 
paper may occur at any sign of $\Omega'$. The instability can operate 
even in a very strong magnetic field which entirely suppresses the 
standard magnetorotational instability. The growth time of instability 
can be as short as few rotation periods. 
\end{abstract}


\maketitle

\section{Introduction}

Instabilities caused by differential rotation of a magnetized gas
may play an important role in enhancing transport processes in various 
astrophysical bodies and laboratory experiments. It is well known since 
the classical papers by Velikhov \cite{vel59} and Chandrasekhar 
\cite{chandra60} that a differentially rotating flow with a negative 
angular velocity gradient and a weak magnetic field is unstable to the 
magnetorotational instability. This instability has been analyzed in 
detail in the astrophysical context (see \cite{fricke69,ach78,bh91}) 
because it can be responsible for transport of the angular momentum in 
various objects ranging from accretion disks to galaxies. In accretion 
disks, this instability is also well studied by numerical simulations 
in both linear and non-linear regimes. Simulations of this instability 
in accretion disks (see, e.g., \cite{bra95,haw95,mat95}) show that the 
generated turbulence can enhance substantially the angular momentum 
transport. 

Astrophysical applications of the magnetorotational instability rise 
great interest in trying to study this instability in laboratory 
\cite{rued01,ji01,bar02,rrb04}. The experiments, however, are complicated 
because very large rotation rates should be achieved. Recently, Hollerbach
and R\"{u}diger \cite{hol05} argued that the rotation rate can be 
substantially decreased adding an azimuthal field. It is known since the
paper by Tayler \cite{tay73} that an azimuthal field produce a strong 
destabilizing effect and, as a result of this additional destabilization,
the critical Reynolds number in experiment can be reduced.  

On the other hand, the magnetorotational instability is not the only  
instability that operates in differentially rotating magnetized flows. 
For example, even a weak axial dependence of the angular velocity can 
result in a double diffusive instability that is often called the 
Goldreich-Schubert-Fricke instability (e.g., \cite{gold}, \cite{arlt04}). 
Note that many previous stability analyses have adopted the Boussinesq 
approximation, and have therefore neglected the effect of 
compressibility. This is allowed if the magnetic field strength is 
essentially subthermal, and the sound speed is much greater than the 
Alfv\'en velocity, $c_{s} \gg c_{A}$ but often this cannot be realized 
in real astrophysical conditions and in many numerical simulations. An 
attempt to consider the effect of compressibility on the 
magnetorotational instability was undertaken by Blaes and  Balbus 
\cite{bla} in the context of astrophysical disks. The authors considered 
a very simplified case of the wavevector parallel to the rotation axis 
and a vanishing radial magnetic field. As a result, the most interesting 
physics has been lost in this study since only the standard 
magnetorotational instability operates in this simple geometry. 

In this paper, we show that a new instability different from the standard 
magnetorotational instability may occur in a compressible differentially 
rotating magnetized flow. This instability appears for any differential 
rotation and may occurs if the magnetic field has non-vanishing radial 
and azimuthal components. The instability can arise even in a 
sufficiently strong magnetic field that suppresses the magnetorotational 
instability. Stability analysis done in this paper will hopefully prove 
to be a useful guide  understand various numerical simulations that 
explore the nonlinear development of instabilities and their effects on 
the resulting turbulent state of rotating magnetized flows.

\section{Basic equations and dispersion relation}

We work in cylindrical coordinates ($s$, $\varphi$, $z$) with the 
unit vectors ($\vec{e}_{s}$, $\vec{e}_{\varphi}$, $\vec{e}_{z}$).
The equations of compressible MHD read  
\begin{eqnarray}
\dot{\vec{v}} + (\vec{v} \cdot \nabla) \vec{v} = - \frac{\nabla p}{\rho} 
+ \vec{g} + \frac{1}{4 \pi \rho} (\nabla \times \vec{B}) \times \vec{B}, 
\end{eqnarray}
\begin{equation}
\dot{\rho} + \nabla \cdot (\rho \vec{v}) = 0, 
\end{equation}
\begin{equation}
\dot{p} + \vec{v} \cdot \nabla p + \gamma p \nabla \cdot 
\vec{v} = 0,
\end{equation}
\begin{equation}
\dot{\vec{B}} - \nabla \times (\vec{v} \times \vec{B}) + \eta
\nabla \times (\nabla \times \vec{B}) = 0,
\end{equation}
\begin{equation}
\nabla \cdot \vec{B} = 0. 
\end{equation} 
Our notation is as follows: $\rho$ and $\vec{v}$ are the density and 
fluid velocity, respectively; $p$ is the gas pressure; $\vec{g}$ is 
gravity that can be important in astrophysical applications; $\vec{B}$ 
is the magnetic field, $\eta$ is the magnetic diffusivity, and $\gamma$ 
is the adiabatic index. For the sake of simplicity, the flow is assumed 
to be isothermal.

The basic state on which the stability analysis is performed is assumed 
to be quasi-stationary with the angular velocity $\Omega = \Omega(s)$ 
and $\vec{B} \neq 0$. Generally, a quasi-stationary basic state cannot 
be achieved for any differentially rotating magnetic configuration, 
therefore we discuss in more detail when this assumption can be 
satisfied. We assume that gas is in hydrostatic equilibrium in the 
basic state, then
\begin{equation}
\frac{\nabla p}{\rho} = \vec{D} + \frac{1}{4 \pi \rho} 
(\nabla \times \vec{B}) \times \vec{B} \;\;, \;\;\;\;
\vec{D} = \vec{g} + \Omega^{2} \vec{s}.
\end{equation}
Generally, a geometry of the magnetic field can be rather complex, and 
our study is primerily addressed such complex magnetic configurations 
where the radial and azimuthal field components are presented. The 
presence of a radial magnetic field and differential rotation in the 
basic state can lead to the development of the azimuthal field. 
Nevertheless, the basic state can be considered in some cases as 
quasi-stationary despite the development of the toroidal field.

For example, if the magnetic Reynolds number is large ($\eta$ is small), 
then one can obtain from Eq.~(4) that the azimuthal field grows 
approximately linearly with time,
\begin{equation}
B_{\varphi}(t) = B_{\varphi}(0) + s \Omega' B_{s} t,
\end{equation} 
where $\Omega' = d \Omega/ds$, and $B_{\varphi}(0)$ is the azimuthal 
field at $t=0$. As long as the second term on the r.h.s. is small
compared to the first one, and
\begin{equation}
t \ll \tau_{\varphi} = \frac{1}{s \Omega'} \; 
\frac{B_{\varphi}(0)}{B_{s}},
\end{equation}
stretching of the azimuthal field does not affect significantly the
basic state; $\tau_{\varphi}$ is the characteristic timescale of 
generation of $B_{\varphi}$. As a result, the basic state can be 
treated as quasi-stationary during the time $t \ll \tau_{\varphi}$. 
If $B_{\varphi}(0)/ B_{s} \gg 1$, then steady-state can be maintained 
during a relatively long time before the generated azimuthal field 
begins to influence the basic state. We will show that the growth 
time of instability can be shorter than $\tau_{\varphi}$ in many cases
of interest.      

If the magnetic Reynolds number is moderate, then stretching of the 
azimuthal field from $B_{s}$ by differential rotation can be compensated 
by ohmic dissipation, and the basic state can be quasi-stationary as 
well. Then, we have from Eq.~(4) the following condition of steady-state
\begin{equation}
[\nabla \times (\vec{v} \times \vec{B})]_{\varphi} =
\eta [\nabla \times (\nabla \times \vec{B})]_{\varphi}, 
\end{equation}    
or
\begin{equation}
\left( \Delta  - \frac{1}{s^{2}} \right) B_{\varphi} = 
- \frac{s}{\eta} \Omega' B_{s}.
\end{equation}
The generated toroidal field is typicaly stronger than the radial 
field by a factor of the order of the magnetic Reynolds number. This 
simple model applies only in the case of moderate Reynolds number 
since the generation of a very strong toroidal field could lead to 
instabilities of the basic state caused, for example, by magnetic 
buoyancy or reconnection. Note that a quasi-stationary basic state 
with non-vanishing radial and azimuthal field components can be achieved 
in other models as well. For example, if the angular velocity depends on 
both the $s$- and $z$-coordinates, then changes in $B_{\varphi}$ caused 
by stretching from the radial and vertical field components due to 
radial and verical shear, respectively, can balance each other in such 
a way that $B_{\varphi}$ will be steady-state. In fact, there is no 
principle difference for instability which mechanism is responsible for 
maintaining a quasi-stationary basic configuration. The only important 
point for our model is the pesence of the magnetic field with 
non-vanishing radial and azimuthal components, but such magnetic 
configurations are rather common in astrophysics (galactic and accretion 
disks, stellar radiative zones, oceans of accreting neutron stars, etc.)

We consider the stability of axisymmetric short wavelength perturbations
with the spacetime dependence $\propto \exp ( \sigma t - i \vec{k}
\cdot \vec{r})$. where $\vec{k}= (k_{s}, 0, k_{z})$ is the wavevector,
$|\vec{k} \cdot \vec{r}| \gg 1$. Small perturbations will be indicated 
by subscript 1, while unperturbed quantities will have no subscript. 
Then, to the lowest order in $|\vec{k} \cdot \vec{r}|^{-1}$ the 
linearized MHD-equations read  
\begin{eqnarray}
\sigma \vec{v}_{1} + 2 \vec{\Omega} \times \vec{v}_{1}
+ \vec{e}_{\varphi} s \Omega' v_{1s}    
= \frac{i \vec{k} p_{1}}{\rho}  
\nonumber \\
- \frac{i}{4 \pi \rho} (\vec{k} \times \vec{B}_{1}) \times \vec{B} ,  
\end{eqnarray}
\begin{equation}
\sigma \rho_{1} - i \rho (\vec{k} \cdot \vec{v}_{1}) = 0, 
\end{equation}
\begin{equation}
\sigma p_{1} -i \gamma p (\vec{k} \cdot \vec{v}_{1}) = 0, 
\end{equation}
\begin{equation}
\sigma \vec{B}_{1} = \vec{e}_{\varphi} s \Omega' B_{1s }  
-i (\vec{B} \cdot \vec{k}) \vec{v}_{1} + i \vec{B} (\vec{k} \cdot 
\vec{v}_{1}), 
\end{equation}
\begin{equation}
\vec{k} \cdot \vec{B}_{1} = 0.
\end{equation}
We neglect ohmic dissipation in the induction equation because the
inverse ohmic decay timescale is small for many cases of interest in
astrophysics.

Generally, the dispersion relation for Eqs.(11)-(15) is rather complicated
and, in this paper, we consider only a particular case when the 
wavevector of perturbations is perpendicular to $\vec{B}$, $\vec{k} 
\cdot \vec{B} =0$. This case, being mathematically much simpler, 
illustrates very well the main qualitative features of the new magnetic 
shear-driven instability. Besides, the standard magnetorotational 
instability does not operate in this case because its growth rate is 
proportional to $\vec{k} \cdot \vec{B}$. Therefore, the difference 
between instabilities is seen most clearly if $\vec{k} \cdot \vec{B}=0$. 

In the case $\vec{k} \cdot \vec{B}=0$, Eqs.~(11)-(15) may be combined 
after some algebra into a fifth-order dispersion relation,
\begin{equation}
\sigma^{5} + \sigma^{3} (\omega^{2}_{0} + \Omega^{2}_{e})
+ \sigma^{2} \omega^{3}_{B \Omega} + \sigma \mu \Omega^{2}_{e} 
\omega^{2}_{0} + \mu \Omega^{2}_{e} \omega^{3}_{B \Omega} = 0
\end{equation}
where we denote
\begin{eqnarray}
\Omega^{2}_{e} = 2 \Omega
(2 \Omega + s \Omega') \;, \;\;
\omega^{2}_{0}= k^{2} ( c^{2}_{s} + c^{2}_{m})\; ,\;\;
\mu = k^{2}_{z} / k^{2}\;,
\nonumber\\
c^{2}_{m} = \frac{B^{2}}{4 \pi \rho} \;, \;\; 
c^{2}_{s} = \frac{\gamma p}{\rho} \;, \;\; \omega^{3}_{B \Omega}=
\frac{k^{2} B_{\varphi} B_{s} s \Omega'}{4 \pi \rho} \;,
\nonumber 
\end{eqnarray}
This equation describes five non-trivial modes that exist in a
rotating magnetized flow if $\vec{k} \cdot \vec{B} =0$. 

In the non-magnetic case, $\vec{B}=0$, Eq.~(16) yields 
\begin{equation}
\sigma^{4} + (\omega_{s}^{2} + \Omega_{e}^{2}) \sigma^{2} +
\mu \omega_{s}^{2} \Omega_{e}^{2} =0,
\end{equation}
where $\omega_{s}= k c_{s}$ is the frequency of sound waves. The 
solution is
\begin{equation}
\sigma_{1,2}^{2} = - \frac{1}{2} (\omega_{s}^{2} + \Omega_{e}^{2})
\pm \sqrt{ \frac{1}{4} (\omega_{s}^{2} + \Omega_{e}^{2})^{2} -
\mu \omega_{s}^{2} \Omega_{e}^{2}}.
\end{equation}
Instability arises only if the well-known Rayleigh criterion is fulfilled,
$\Omega_{e}^{2}< 0$. In this case, the inertial mode is unstable which
corresponds to the upper sign. The sound mode that corresponds to the
lower sign is always stable.

To have an idea about the properties of dispersion equation 
(16), we can consider a particular case of flow with $\Omega \propto 
s^{-2}$. Then, $\Omega^{2}_{e}=0$ and we have from Eq.~(16)
\begin{equation}
\sigma^{3} + \sigma \omega^{2}_{0} + \omega^{3}_{B \Omega}=0.
\end{equation} 
The solutions of this equation are
\begin{equation}
\sigma_{1} = u+v \;, \;\; \sigma_{2,3} = - \frac{1}{2} (u+v)
\pm \frac{i \sqrt{3}}{2} (u - v),
\end{equation}
where
\begin{equation}
(u, v) = \left( - \frac{\omega^{3}_{B \Omega}}{2} \pm 
\sqrt{ \frac{\omega^{6}_{B \Omega}}{4} + \frac{\omega^{6}_{0}}{27}} 
\right)^{1/3}.
\end{equation}
At least, one of the roots has a positive real part (instability) if
$u+v \neq 0$. The latter condition is equivalent $\omega^{3}_{B \Omega}
\neq 0$ that is the criterion of instability in this simple case.
It is clear from this simple example that the quantity 
$\omega_{B \Omega}$ plays a crucial role for stability of magnetized
compressible flows.

\section{Criteria and growth rate of instability}

The conditions under which Eq.~(16) has unstable solutions can be 
obtained by making use of the Routh-Hurwitz theorem (see \cite{hen}, 
\cite{alek}). In the case of the dispersion equation of a fifth order, 
the Routh-Hurwitz criteria are written, for example, in \cite{urp}.
According to these criteria, Eq.~(16) has unstable solutions if one 
of the following inequalities is fulfilled
\begin{equation}
\mu \Omega^{2}_{e} \omega^{3}_{B \Omega} < 0 \;,\;\;
\omega^{3}_{B \Omega} > 0 \;,\;\;
(\omega^{3}_{B \Omega})^{2}  < 0.
\end{equation}
These inequalities yield the criterion of instability
\begin{equation}
\omega^{3}_{B \Omega} \neq 0.
\end{equation}
Apart from differential rotation, this criterion requires non-vanishing 
radial and azimuthal field components. The vertical component of 
$\vec{B}$ is unimportant for criterion (23), and the instability may
occur even in a plane parallel magnetic field with components only in
radius and azimuth. The direction of $\vec{B}$ and the sign of $\Omega'$ 
are insignificant, and the instability may occur for both the inward and 
outward decreasing angular velocity. Note that this is in contrast with 
the magnetorotational instability that can arise only if $\Omega' < 0$. 
Another important difference is that the magnetorotational instability 
is suppressed by a sufficiently strong field, whereas the instability 
given by Eq.~(23) can arise even in a very strong field.

To calculate the growth rate in the general case it is convenient to 
introduce dimensionless quantities
\begin{equation}
\Gamma= \frac{\sigma}{\Omega_{e}} \;,\;\; \xi = 
\frac{1}{x^{2}} \frac{\omega^{2}_{0}}{\Omega^{2}_{e}} \;,\;\; 
\zeta = \frac{1}{x^{2}} \frac{\omega^{3}_{B \Omega}}{\Omega^{3}_{e}}
\;, \;\; x=ks
\end{equation}
(we assume that $\Omega^{2}_{e} >0$). Note that the parameters 
$\xi$ and $\zeta$ do not depend on the wavevector. Then, Eq.~(16) 
becomes 
\begin{equation}
\Gamma^{5} + \Gamma^{3} (1 + \xi x^{2}) + \Gamma^{2} \zeta x^{2} + 
\Gamma \mu \xi x^{2} + \mu \zeta x^{2} = 0.
\end{equation}
This equation was solved numerically for different $\mu$, $\xi$, and
$\zeta$  by computing the eigenvalues of the matrix whose characteristic 
polynomial is given by Eq.~(16) (see \cite{press}, for details).

In Fig.~1, we plot the dependence of the real and imaginary parts of 
$\Gamma$ on $x$ for $\mu =0.3$, $\xi =0.1$ and $\zeta=0.1$. The solid 
lines show the growth rate and frequency for complex conjugate roots, and 
the dashed line for a real root. As  mentioned, there should be no 
instability in the incompressible limit because  all the considered 
perturbations are stable with respect to the standard magnetorotational 
instability. Our calculations, however, clearly indicate that some roots 
have a positive real part and, hence, there should exist a new 
shear-driven instability in the compressible flow with $\zeta \neq 0$. 
There are two pairs of unstable complex conjugate roots and one real 
stable root with negative Re $\Gamma$. In the considered domain of 
parameters, Im $\Gamma$ for complex roots is typically $\sim 10-30$ 
times greater than Re $\Gamma$ except the region of not very large 
$x^{2} \sim 10-50$ where they are of the same order of magnitude (but 
still Im $\Gamma >$ Re $\Gamma$). One pair of unstable roots has a very 
small growth rate $\sim 10^{-4} \Omega_{e}$, but another one grows much 
faster. For these roots, the growth rate is $\approx 0.5 \Omega_{e}$ 
and varies very slowly with the wavelength of perturbations. Note that 
calculations for other values of the parameters show that typically 
Re $\Gamma \sim 0.5$ if $\xi \sim \zeta$, but Re $\Gamma$ becomes 
smaller if $\xi \gg \zeta$. This is qualitatively clear because the 
case $\xi \gg \zeta$ corresponds to the incompressible limit when the 
considered instability is substantially suppressed. 

In Fig.~2, we plot the dependence of Re $\Gamma$ on $x$ for the case 
$\mu=0.3$,  $\xi=0.1$  and $\zeta= -0.1$. We do not plot Im $\Gamma$ 
because this dependence does not differ much from what is shown in 
Fig.~1. The change of sign alters qualitatively the behavior of roots. 
If $\zeta$ is negative then all oscillatory modes are stable (Re $\Gamma 
< 0$) but the real mode becomes unstable. This conclusion is completely 
consistent with our analytical consideration of Eq.~(14). It is worth 
mentioning that calculations for other $\mu$, $\xi$, and $\zeta$ also 
indicate that this sort of behavior is rather general, and the 
non-oscillatory mode is typically unstable for negative $\zeta$ whereas 
the oscillatory modes are unstable for positive $\zeta$. Like the 
previous case, the growth rate weakly depends on the wavelength except 
the region $x^{2} < 200$ where this dependence is stronger. The 
characteristic value of the growth rate is larger for negative $\zeta$ 
and reaches $\approx \Omega_{e}$ for $x^{2} \geq 200$. Note that this is 
typical also for other values of $\mu$ and $\xi$ and that a 
non-oscillatory mode (negative $\zeta$) grows faster than oscillatory 
modes (positive $\zeta$) for the same $|\zeta|$. 

\begin{figure}
\includegraphics[width=9cm]{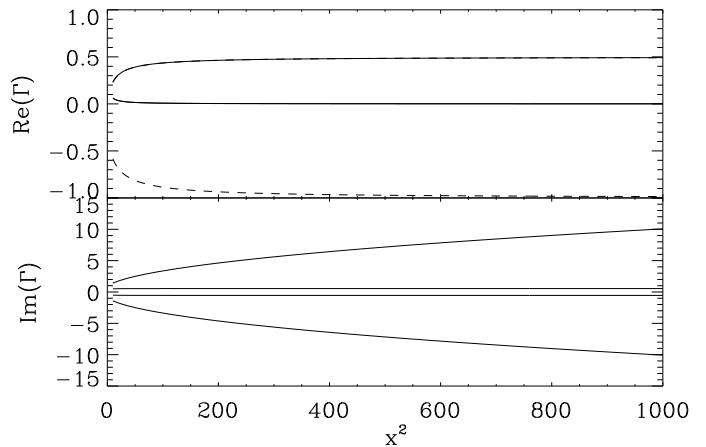}
\caption{The dependence of the real and imaginary parts of $\Gamma$ on 
$x^{2}$ for $\mu =0.3$, $\xi = 0.1$, and $\zeta=0.1$. Solids lines show 
the growth rate and frequency of complex roots, and the dashed line 
corresponds to the real root.}
\end{figure}

\begin{figure}
\includegraphics[width=9.0cm]{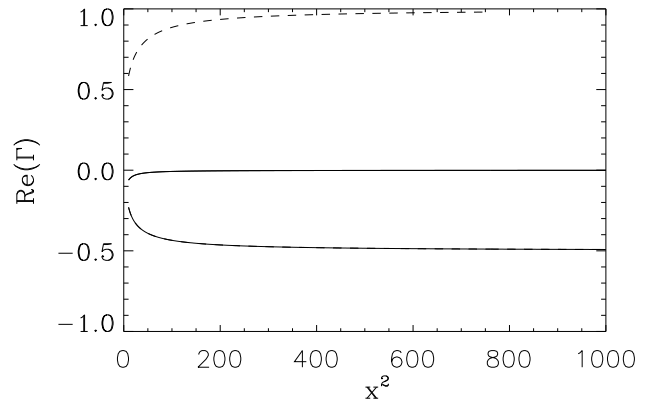}
\caption{The dependence of the growth rate on $x^{2}$ for $\mu=0.3$, 
$\xi = 0.1$, and $\zeta=-0.1$. The solid and dashed lines correspond to 
complex and real roots, respectively.}
\end{figure}

Fig.~3 illustrates the behavior of roots as functions of the parameter
$\zeta$ for fixed value of $x$. It is seen that Re $\Gamma$ vanishes for 
both oscillatory and non-oscillatory modes when $\zeta$ goes to zero. 
Since $\zeta \propto \omega_{B \Omega}$, the instability occurs only if 
$\omega_{B \Omega} \neq 0$ in complete agreement with the criterion (13). 
As usual, the real root is positive (instability) at $\zeta < 0$ whereas 
the oscillatory roots have positive real parts at $\zeta > 0$. For the 
same $|\zeta|$, the growth rate is larger for negative $\zeta$. 

\begin{figure}
\includegraphics[width=9.0cm]{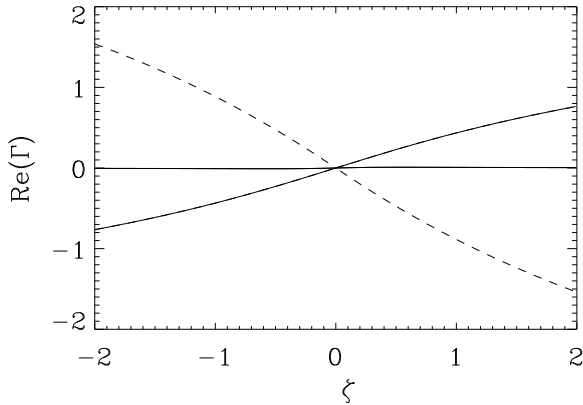}
\caption{The dependence of Re $\Gamma$ on $\zeta $ for $\mu=0.3$, 
$x^2=10$ and $\xi = 1$. Solid and dashed lines correspond to complex 
and real roots, respectively.}
\end{figure}

\section{Discussion}

To summarize then, we have considered the instability caused by 
differential rotation of compressible magnetized gas. To illustrate the 
main qualitative features of the instability associated to 
compressibility and shear, we analyzed a particular case of perturbations 
with the wavevector $\vec{k}$ perpendicular to the magnetic field 
$\vec{B}$. In this case, the standard magnetorotational instability, 
well-studied in incompressible fluids (see, e.g., \cite{vel59}, 
\cite{chandra60}, \cite{bh91}), does not occur because its growth rate is 
proportional to $(\vec{k} \cdot \vec{B})$. Nevertheless, even 
perturbations with $\vec{k} \cdot \vec{B} =0$ turn out to be instable if 
the necessary condition of the new instability, $\zeta \propto B_{s} 
B_{\varphi} \Omega' \neq 0$, is satisfied.

In our stability analysis, we assume that the basic state is 
quasi-stationary. This assumption can be fulfilled in many cases of 
astrophysical interest despite the development of the azimuthal field 
from the radial one due to differential rotation. For instance, if the 
magnetic Reynolds number is large then the timescale of generation of 
the toroidal field is $\sim \tau_{\varphi}$ if $B_{\varphi}(0) > B_{s}$. 
Instability can be considered in a quasi-stationary approximation if 
its growth time is shorter than $\tau_{\varphi}$. As it is seen from 
Eq.~(21), the growth rate of instability in the case of a strong 
compressibility can be roughly estimated as $\omega_{B \Omega}$. Then, 
the condition of quasi-stationarity reads $\omega_{B \Omega} \gg 
1/\tau_{\varphi}$, or
\begin{equation}
k c_{As} > s \Omega' \left( \frac{B_{s}}{B_{\varphi}(0)} \right)^{2}.
\end{equation} 
Since the l.h.s. of this equation is proportional to $k$ but the r.h.s. 
does not depend on $k$, there always exists the range of $k$ for which 
Eq.~(26) can be satisfied and the basic state is quasi-stationary.

The considered instability is related basically to shear and 
compressibility of a magnetized gas. In the incompressible limit that 
corresponds to $c_{s} \rightarrow \infty$ or $\omega_{0}^{2} \rightarrow 
\infty$, we have from Eq.~(16)
\begin{equation}
\sigma (\sigma^{2} + \mu \Omega_{e}^{2}) =0,
\end{equation}
and the instability disappears. This is a principle difference to other 
well-known instabilities caused by differential rotation such as the 
Rayleigh or magnetorotational instabilities. Note that an attempt to
consider instability associated to compressibility of differentially
rotating magnetized gas has been undertaken by Blaes \& Balbus 
\cite{bla}. These authors, however, analyzed only the unperturbed 
configuration where the magnetic field has vertical or azimuthal 
components, but such configurations are stable in accordance to our 
criterion (23).

The properties of the considered instability are very much different 
from those of other instabilities which can occur in cylindrical 
magnetized flows. The necessary condition of the new instability (23) 
can be satisfied for both outward increasing and decreasing $\Omega(s)$, 
whereas the magnetorotational instability occurs only if $\Omega(s)$ 
decreases with $s$. The found instability operates only if the basic 
magnetic configuration is relatively complex with non-vanishing radial 
and azimuthal field components while the standard magnetorotational 
instability can arise also if both these components are vanishing and 
only $B_{z} \neq 0$. 

This new  instability can be either oscillatory or non-oscillatory, 
depending on the sign of $\zeta$, whereas the standard magnetorotational 
instability is always non-oscillatory. Typically, the considered 
instability is non-oscillatory if $\zeta < 0$ and oscillatory if $\zeta 
>0$. One more important difference is associated with the dependence on 
the magnetic field strength. A sufficiently strong magnetic field, 
satisfying the inequality $(\vec{k} \cdot \vec{B})^{2} > 8 \pi \rho 
s \Omega |\Omega'| (k^{2}_{z}/k^{2})|$, completely suppresses the 
standard magnetorotational instability. On the contrary, the instability 
discovered in our study cannot be suppressed even in very strong magnetic 
fields as it is seen from the criterion (23). All this comparison allows 
us to claim that our analysis demonstrates the presence of the new 
instability in compressible cylindrical flow.
 
The growth rate of the newly found instability can be rather large and 
reach $\sim \Omega_{e}$. Basically, the growth rate is larger for 
non-oscillatory modes which are unstable if $\omega^{3}_{B \Omega} < 0$. 
The growth rate depends on compressibility,  being smaller for a low 
compressibility. The incompressible limit (Boussinesq approximation) 
corresponds to $c_{s} \gg c_{A}$, and the considered instability is 
inefficient in this limit because of a low growth rate. However, in the 
case of a strong field with $c_{A} \sim c_{s}$ when the Boussinesq 
approximation does not apply, the instability can be much more efficient 
than the magnetorotational instability.

\vspace{0.5cm}
\noindent
{\it Acknowledgments.}
VU thanks INAF-Ossevatorio Astrofisico di Catania for hospitality and
financial support under the Marie Curie Senior Research fellowship 
(contract MTKD-2004-002995). We thank L.Santagati for careful reading 
of the manuscript.

\end{document}